\documentclass[runningheads]{llncs}
\usepackage{graphicx}
\usepackage{bbm}
\usepackage{booktabs}
\usepackage{multirow}

\usepackage{amsmath,amsfonts,bm}

\def\eqref#1{equation~\ref{#1}}

\def\1{\bm{1}}

\DeclareMathAlphabet{\mathsfit}{\encodingdefault}{\sfdefault}{m}{sl}
\SetMathAlphabet{\mathsfit}{bold}{\encodingdefault}{\sfdefault}{bx}{n}

\begin{document}

\title{A Multimodal Machine Learning Framework for Teacher Vocal Delivery Evaluation}
%
\titlerunning{ML for Teacher Vocal Delivery Evaluation}
%

\author{Hang Li \and Yu Kang \and Yang Hao \and Wenbiao Ding \and Zhongqin Wu \and Zitao Liu\thanks{Corresponding Author: Zitao Liu}}

\authorrunning{H. Li et al.}
%

\institute{TAL Education Group, Beijing, China \\
\email{\{lihang4,kangyu,haoyang2,dingwenbiao,wuzhongqin,liuzitao\}@tal.com}}

\maketitle              
\begin{abstract}

The quality of vocal delivery is one of the key indicators for evaluating teacher enthusiasm, which has been widely accepted to be connected to the overall course qualities. However, existing evaluation for vocal delivery is mainly conducted with manual ratings, which faces two core challenges: subjectivity and time-consuming. In this paper, we present a novel machine learning approach that utilizes pairwise comparisons and a multimodal orthogonal fusing algorithm to generate large-scale objective evaluation results of the teacher vocal delivery in terms of fluency and passion. We collect two datasets from real-world education scenarios and the experiment results demonstrate the effectiveness of our algorithm. To encourage reproducible results, we make our code public available at \url{https://github.com/tal-ai/ML4VocalDelivery.git}.

\end{abstract}
\section{Introduction}
Teacher enthusiasm has been widely accepted by recent researches that is highly correlated with the high-quality instructions, which provides students with learning opportunity and fosters their learning and achievement \cite{feldman2007identifying,moulding2010intelligent,zeidner2007test,liu2020dolphin}. To evaluate teacher enthusiasm, multiple statistical algorithms focusing on counting and scoring different aspects of instruction behaviors, i.e., vocal delivery, facial expressions, have been employed as the basic indicators of enthusiastic teaching in their own systems \cite{bettencourt1983effects,brigham1992teacher,collins1978effects,murray1983low,li2020siamese}. Among these studies, vocal delivery is one of the most commonly accepted indicators due to its irreplaceability in student-teacher communication. Therefore, we focus on improving the existing vocal delivery evaluation (VDE) via the advanced machine learning algorithms.

Traditionally, VDE is conducted by human observers and the evaluation results face two challenges: (1) \emph{subjectivity}: human annotators may have different understandings about evaluation rules; and (2) \emph{time-consuming}: vocal delivery manual evaluation requires annotators to examine the vocal samples multiple times. To solve these two challenges, we propose a multimodal machine learning framework to conduct objective VDE in terms of both fluency and passion. The fluency indicator is designed to detect poor articulations between the words and topics, and the passion indicator is utilized to evaluate the variations of pitch, volume, and speed. 

In summary, the contributions of this work are: (1) we alleviate the subjectivity problem in current VDE by utilizing the pairwise comparisons; (2) we propose a multimodal orthogonal fusing algorithm, which helps embeddings from different unimodal pre-trained models fuse in an informative way; (3) we demonstrate that our proposed method is able to provide accurate and objective evaluation results.

\section{Label Generation via Pairwise Comparison}
\label{sec:comparison}
In our framework, in order to obtain reliable training labels for VDE, we design a two-step label generation algorithm via pairwise comparison to eliminate the discrepancies caused by the ambiguous descriptions to the anchors of some subjective perceptions such as passion \cite{bradley1952rank,maystre2015fast,plackett1975analysis}. 

\noindent{\bf Anchor Selection} We collect a moderate-size unlabeled dataset $S=\{s_i\}_{i=1}^N$ via uniform sampling. After that, for each paired samples $(s_i,s_j)$, we ask human annotators to judge which sample is better fitting the requirements (e.g., ``Is sample A more passionate than sample B?"). After collecting plenty of these comparing results, we model the probability of choosing $s_i$ over $s_j$ by utilizing the Bradley-Terry model \cite{bradley1952rank}, i.e., $P(s_i>s_j) = f(a_i-a_j)$ where $f(u)=\frac{1}{1+\exp(-u/\sigma)}$, $a_i$ is the estimated ranking score and $\sigma$ is standard deviation of $A=\{a_i\}_{i=1}^N$. Following the prior work by Tsukida and Gupta \cite{tsukida2011analyze}, the ranking scores $A$ is obtained through maximum a posteriori estimation. After that, we carefully choose $L$ anchor samples $\mathcal{G} = \{s_{g_1}, \cdots, s_{g_L}\}$ by percentiles that represent the ranking score distribution.

\noindent{\bf Comparison Labeling} Once we obtain the anchor samples, we label the remaining samples based on their  comparing results with selected anchors. More specifically, for a new sample $s^*$, we first conduct its pairwise comparisons with each anchor in $\mathcal{G}$. Then, similar to the ranking score generation process in the anchor selection step, we learn the Bradley-Terry model from these comparison results and obtain the corresponding ranking score $a^*$. Finally, we compare $a^*$ with the ranking scores, i.e., $\{a_{g_1}, \cdots, a_{g_L}\}$ of our select anchors in $\mathcal{G}$ and the final label $y^*$ is obtained by computing number of anchors ordered after $s^*$, i.e., $y^* = \sum^{L}_{l=1}\mathbbm{1}_{a^*>a_{g_l}}$, where $\mathbbm{1}_{(\cdot)}$ is an indicator function.


\section{Multimodal Learning}
\label{sec:multimodal}
\label{sec:method}
\begin{figure*}[!hbpt]
    \vspace{-0.2cm}
    \begin{center}
    \includegraphics[width=0.85\textwidth]{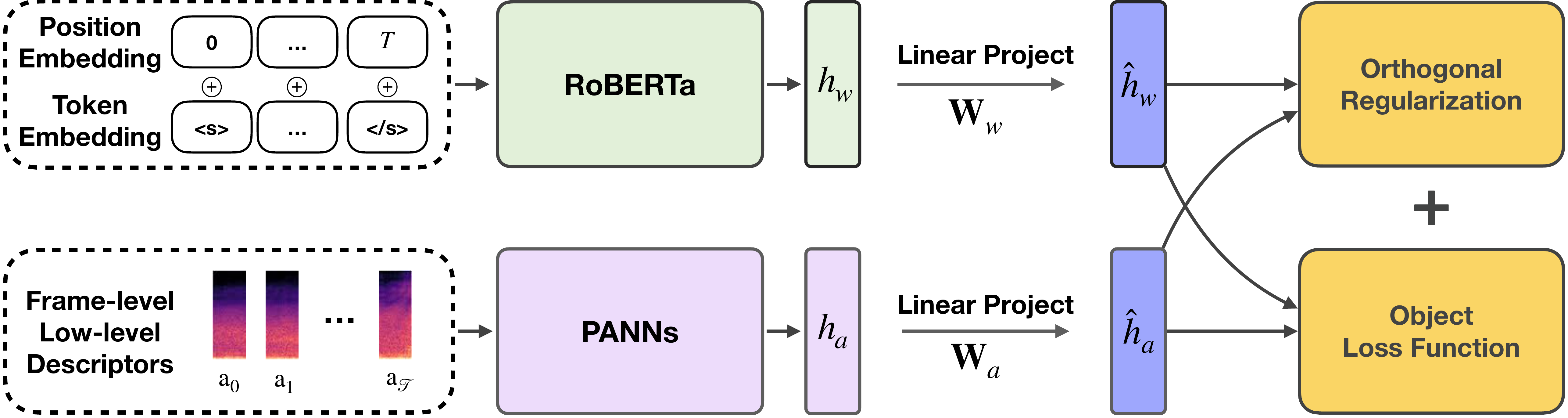}
    \end{center}
    \vspace{-0.5cm}
    \caption{The proposed multimodal learning framework.}
    \label{fig:model}
    \vspace{-0.5cm}
\end{figure*}

The traditional evaluation of vocal delivery usually involves complicated considerations on multiple facets of speeches \cite{bettencourt1983effects,brigham1992teacher,collins1978effects}. To make full use of these information in each speech sample, we propose a multimodal learning framework with three modules: a language encoder, an audio encoder, and a multimodal fusion block. The overall framework architecture is shown in Fig. \ref{fig:model}.

\noindent{\bf Language Encoder} The pre-trained language models like BERT \cite{devlin2018bert}, RoBERTa \cite{liu2019roberta}, BART \cite{lewis2019bart} have been demonstrated to have strong capabilities in capturing semantic information. In our framework, we choose to use RoBERTa as our backbone model which accepts text token embeddings combining with their corresponding position embeddings as inputs. Following prior researches \cite{liu2019roberta}, we use the first token's output representation $\mathbf{h}_w$ as the extracted semantic sentence embedding.

\noindent{\bf Audio Encoder} Similar to language encoder, we use a pretrained audio neural networks (PAANs) \cite{kong2020panns} as our backbone module to extract acoustic features. The inputs of the audio encoder are the frame-level low-level descriptors and the output is a single vector $\mathbf{h}_a$, which summarizes the acoustic features of the entire utterance.


\noindent{\bf Orthogonal Fusion} Multimodal learning aims to exhibit and capture information from different modalities and therefore, we propose an orthogonal fusion method to enforce representations from different modalities to be dissimilar. Specifically, we design an additional orthogonal regularization penalty as follows: $\mathcal{L}_{\mathrm{Orth}} = \frac{|{(\mathbf{W}_a \cdot \mathbf{h}_{a})}^\top (\mathbf{W}_w \cdot \mathbf{h}_{w})|}{\left \| \mathbf{W}_a\cdot\mathbf{h}_{a} \right \|  \left \| \mathbf{W}_w\cdot\mathbf{h}_{w} \right \| }$, where $\mathbf{W}_w$ and $\mathbf{W}_a$ are trainable parameters that project $\mathbf{h}_{w}$ and $\mathbf{h}_{a}$ to the same hidden space respectively. In the final objective functions, we use the fused representation $\mathbf{h}^{fuse}$, i.e., $\mathbf{h}^{fuse} = \mathbf{W}_w \cdot \mathbf{h}_{w} \ \oplus \ \mathbf{W}_a \cdot \mathbf{h}_{a}$ to optimize the VDE loss together with the regularization term $\mathcal{L}_{\mathrm{Orth}}$.




\section{Experiments}
We evaluate teacher vocal delivery in two aspects: fluency and passion. We collect two datasets from real-world K-12 education scenarios: (1) the \emph{Passion} dataset contains 18,000 teacher speech samples extracted from a third-party online class platform; and (2) the \emph{Fluency} dataset includes 15,000 utterances and each sample is labeled based on its fluency level. The sample labels for these two datasets are obtained through pairwise comparisons discussed in Section \ref{sec:comparison}. We choose two anchors, i.e., set $L = 2$, which represent the 25\% and 75\% percentiles. Hence, samples are split into three groups: high, medium and low. In terms of model training, we exclude samples of medium group to reduce the ambiguity. 1,000 utterances are randomly sampled from each dataset and used as test data. Additionally, we perform a 20\%/80\% split over the remaining dataset to generate validation and train sets. We choose to use accuracy and macro F1-score as our evaluation metrics.

To validate the pairwise-comparing algorithm, we ask three experts to justify them. From the results, we find more than 95\% of these positive and negative labeled samples are accepted by at least two experts. To assess the effectiveness of our approach, we carefully choose the following methods as our baselines: (1) $\emph{RoBERTa}$: a strong large-scale pre-trained language model only uses text as input. (2) $\emph{PANNs}$: a uni-modal pre-trained model that only uses audio signals as input. (3) $\emph{Concat}$: a multimodal model which uses both pre-trained RoBERTa and PANNs to extract features and simply concatenates the representations of different modalities for classification. The detailed results for both fluency and passion datasets are shown in Table~\ref{tab:result}.

From Table~\ref{tab:result}, we have several observations: (1) by comparing \emph{RoBERTa} and \emph{PANNs} on Fluency dataset, we find language information is more important than audio for fluency evaluation; (2) we observe \emph{PANNs} outperforms \emph{RoBERTa} on Passion dataset, which is consistent with our expectation that acoustic features should be better in evaluating the passion of the utterance; (3) when comparing \emph{Concat} with prior two unimodal models, we find it outperforms the two unimodal baselines by a great margin, which indicates the effectiveness of multimodal learning; (4) by comparing \emph{Ours} to \emph{Concat}, we find the model's performance is further improved.

\begin{table}[!hbpt]
    \caption{Model performances on two datasets. $\mathrm{Acc}$ and $\mathrm{F1_{macro}}$ indicate the accuracy and macro F1-score respectively.}
    \label{tab:result}
    \centering
    \begin{tabular}{@{}ccccccccc@{}} \toprule
    \multirow{2}{*}{Task} & \multicolumn{2}{c}{PAANs} & \multicolumn{2}{c}{RoBERTa} & \multicolumn{2}{c}{Concat} & \multicolumn{2}{c}{Ours} \\ \cmidrule(l){2-9} 
     & Acc & $\mathrm{F1_{macro}}$ & Acc & $\mathrm{F1_{macro}}$ & Acc & $\mathrm{F1_{macro}}$ & Acc & $\mathrm{F1_{macro}}$ \\ \midrule
    \ \ \ Passion \ \ \ & 0.775 & 0.723 & 0.763 & 0.714 & 0.808 & 0.758 & 0.846 & 0.805 \\
    \ \ \ Fluency \ \ \ & 0.654 & 0.628 & 0.788 & 0.777 & 0.838 & 0.828 & 0.872 & 0.862 \\ \bottomrule
    \end{tabular}
\end{table}

\section{Conclusion}
In this work, we present an efficient machine learning approach to evaluate teacher vocal delivery for online classes. Experiments demonstrate that our framework achieves accurate evaluations in terms of both fluency and passion aspects. In the future, we would like to conduct further researches to the other facets of the teacher enthusiasm.


\section*{Acknowledgments}

This work was supported in part by National Key R\&D Program of China, under Grant No. 2020AAA0104500 and in part by Beijing Nova Program (Z201100006820068) from Beijing Municipal Science \& Technology Commission.

\bibliographystyle{splncs04}
\bibliography{aied2021}

\end{document}